# Optimum Decoder for an Additive Video Watermarking with Laplacian Noise in H.264

Nematollah Zarmehi, Morteza Banagar
School of Electrical and Computer Engineering
Faculty of Engineering, University of Tehran
Tehran, Iran
zarmehi@ut.ac.ir, morteza.banagar@ut.ac.ir

Mohammad Ali Akhaee
School of Electrical and Computer Engineering
Faculty of Engineering, University of Tehran
Tehran, Iran
akhaee@ut.ac.ir

*Abstract*—In this paper, we investigate an additive video watermarking method in H.264 standard in presence of the Laplacian noise. In some applications, due to the loss of some pixels or a region of a frame, we resort to Laplacian noise rather than Gaussian one. The embedding is performed in the transform domain; while an optimum and a sub-optimum decoder are derived for the proposed Laplacian model. Simulation results show that the proposed watermarking scheme has suitable performance with enough transparency required for watermarking applications.

I. INTRODUCTION

The ongoing growth of digital multimedia and their widespread publications have brought some worries in the field. Prevention from copying or unauthorized publications and the proof of ownership is of vital importance to their manufacturers. For instance, nowadays in the virtual area of the Internet, unauthorized copying and publications can be done easily and in a very short time, which is a problem that needs to be solved.

Watermarking is a solution to the mentioned problem. Adding a watermark to the host signal, which requires a slight change in it, enables us to understand the identity of the host signal. In order to prevent the unauthorized copying felony, one can think of methods to embed and then extract the watermark from a document; if the absence of such watermarks is detected, we can prevent the document from being copied. Watermarking is done in different types of documents – text, voice, image, and video to name some – and has many applications, such as owner identification, proof of ownership, content authentication, copy control, secret communication, and legacy enhancement [1, 2].

One medium for data hiding is the video signal which recently has drawn a lot of attention to itself. Video signals are usually published in compressed format. For the compressed format, some standards have been developed which one of its latest standards is H.264. Video watermarking can be done both in raw and compressed domains. In raw domain, a good watermarking scheme must be robust against compression standards. Technically speaking, when a video watermarking in the raw domain is exposed to both compression and decompression, the watermarked data must still be detectable. But since watermarking in the raw domain does not care about compression standards, it can be done in a simpler manner compared to watermarking in the compressed domain. On the other hand, due to the abundance of compressed format in video signals, most proposed watermarking schemes are in compressed domain. Here, we perform watermarking in compressed domain on H.264 standard.

In this paper, we investigate the additive watermarking method in the presence of Laplacian noise [3]. In most applications, noise is modeled as a Gaussian random variable, but in our analysis, Gaussian distribution cannot be a good model for noise. As an example, corrupted pixels of a region of an image or of a frame of a video are usually modeled with impulsive noise [4], which needs a larger heavy-tail distribution than Gaussian one [5]. This is why the Laplacian distribution is considered in our analysis.

In our process of transmitting the watermarked video, we consider the distribution of noise to be Laplacian. The embedding is performed on the DCT coefficients of $4 \times 4$ macro-blocks of I-frames. In order to analyze the proposed watermarking scheme, a suitable statistical model is needed for the DCT coefficients. A general form applying to DCT coefficients is General Gaussian Distribution (GGD) [6], which is stated as follows:

$$f_X(x) = A e^{-|\beta(x-m)|^c}, \qquad (1)$$

where

$$\beta = \frac{1}{\sigma}\sqrt{\frac{\Gamma(3/c)}{\Gamma(1/c)}} \quad , \quad A = \frac{\beta c}{2\Gamma(1/c)} \qquad (2)$$

and $m$ is the mean, $\sigma$ is the standard deviation of the distribution, $c$ is the shape parameter, and $\Gamma(.)$ is the Gamma function.

Here, for modeling the DCT coefficients, we set $c = 1$, which results in exactly the Laplacian distribution. As mentioned before, the channel noise is modeled with the Laplacian distribution, as well.

## II. WATERMARK EMBEDDING

Based on a pseudorandom key generated at the transmitter side, $N$ high frequency DCT coefficients from a $4 \times 4$ macro-block of I-frames are selected and the watermarked bit is embedded into these coefficients according to the following rule:

$$x'_i = \begin{cases} x_i - a & \text{for embedding } 0 \\ x_i + a & \text{for embedding } 1 \end{cases} ; \quad i = 1, \ldots, N, \quad (3)$$

where $x_i$ is the selected $i^{th}$ coefficient, $a$ is the gain factor, and $x'_i$ is the watermarked coefficient.

## III. WATERMARK DECODING

What is delivered to the receiver side is the watermarked signal plus channel noise.

$$y_i = x'_i + n_i \quad (4)$$

This additive noise is an i.i.d. Laplacian distributed sample with the parameter $\lambda_2$. Samples of the host signal are also i.i.d. Laplacian distributed with parameter $\lambda_1$. We assume the variance of the noise to be $1/g^2$ of the variance of the host signal. Hence, we have:

$$f_x(x_i) = \frac{\lambda_1}{2} \exp(-\lambda_1 |x_i|), \quad (5)$$

and

$$f_n(n_i) = \frac{\lambda_2}{2} \exp(-\lambda_2 |n_i|) \; ; \; \lambda_2 = g\lambda_1. \quad (6)$$

The channel noise is supposed to be independent of the host signal, and thus the joint distribution of the noise and host signal can be written as follows [7]:

$$f_{x_i+n_i}(y_i) = f(y_i) = \frac{1}{2}\frac{\lambda_1}{\lambda_2}\frac{1}{1-(\frac{\lambda_1}{\lambda_2})^2}[\frac{1}{\lambda_1}e^{-\frac{|y_i|}{\lambda_2}} - \frac{1}{\lambda_2}e^{-\frac{|y_i|}{\lambda_1}}]. \quad (7)$$

Consider the following hypothesis test for extracting the watermarked data:

$$\begin{cases} H_0 : 0 \text{ is embedded} \\ H_1 : 1 \text{ is embedded} \end{cases} \quad (8)$$

In order to extract the watermark, the maximum likelihood decoding (MLD) rule is exploited:

$$L(y_1, y_2, \ldots, y_N) = \frac{\prod_{i=1}^{N} f(y_i | w=1)}{\prod_{i=1}^{N} f(y_i | w=0)} \quad (9)$$

$$= \prod_{i=1}^{N} \frac{\exp(-|y_i - a|/g\lambda_1) - (1/g)\exp(-|y_i - a|/\lambda_1)}{\exp(-|y_i + a|/g\lambda_1) - (1/g)\exp(-|y_i + a|/\lambda_1)} \underset{H_0}{\overset{H_1}{\gtrless}} 1.$$

Since the logarithmic function is an increasing function, we can take the logarithm of the latter equation as below:

$$\sum_{i=1}^{N} \ln[\frac{\exp(-|y_i - a|/g\lambda_1) - (1/g)\exp(-|y_i - a|/\lambda_1)}{\exp(-|y_i + a|/g\lambda_1) - (1/g)\exp(-|y_i + a|/\lambda_1)}] \underset{H_0}{\overset{H_1}{\gtrless}} 0 \quad (10)$$

Or in the simpler form as:

$$\sum_{i=1}^{N} \frac{|y_i + a| - |y_i - a|}{2} + \frac{g\lambda_1}{2}\sum_{i=1}^{N} \ln[\frac{1-(1/g)\exp(-|y_i - a|(g-1)/\lambda_1)}{1-(1/g)\exp(-|y_i + a|(g-1)/\lambda_1)}] \underset{H_0}{\overset{H_1}{\gtrless}} 0. \quad (11)$$

The decision rule at the receiver side can be more simplified as follows:

$$D = \sum_{i=1}^{N} l(y_i, a) + \sum_{i=1}^{N} h(y_i, \lambda_1, g, a) \underset{H_0}{\overset{H_1}{\gtrless}} 0, \quad (12)$$

The block diagram of the optimum receiver is shown in Figure 1.

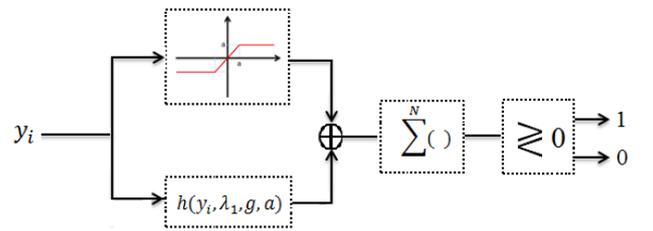

Figure 1. Block diagram of the optimum decoder.

## IV. PERFORMANCE ANALYSIS

In order to analyze the performance of the receiver and computing the error probability, the distribution of $D$ is needed. The complexity of computing this distribution is due to $h(y_i, \lambda_i, g, a)$ which is ignored in the sub-optimum method.

In Figure 2 we illustrated the block diagram of the sub-optimum decoder.

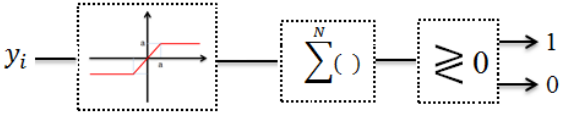

Figure 2. Block diagram of the sub-optimum decoder.

$$Z = \sum_{i=1}^{N} z_i = \sum_{i=1}^{N} l(y_i, a) \underset{H_0}{\overset{H_1}{\gtrless}} 0,$$

$$z_i = l(y_i, a) = \begin{cases} a & y_i > a \\ y_i & -a < y_i < a \\ -a & y_i < -a \end{cases} \quad (13)$$

We calculate the error probability of this sub-optimum receiver and show that the error probability of the optimum receiver is better than the sub-optimum one. To do so, we first derive the conditional density function of the following random variable.

$$z_{i|H_0} = l(x_i - a + n_i, a) = \begin{cases} a & x_i + n_i > 2a \\ x_i + n_i - a & 0 < x_i + n_i < 2a \\ -a & x_i + n_i < 0 \end{cases} \quad (14)$$

The calculations are as follows:

$$p(z_{i|H_0} = a) = p(x_i + n_i > 2a) = \int_{2a}^{+\infty} f_y(u) du$$
$$= \frac{1}{2} \frac{g}{g^2 - 1} [g e^{-2a/g\lambda_1} - (1/g) e^{-2a/\lambda_1}] = P, \quad (15)$$

$$p(z_{i|H_0} = -a) = p(x_i + n_i < 0) = 0.5, \quad (16)$$

$$f_{z_{i|H_0}}(z) = 0.5\delta(z+a) + P\delta(z-a) + f_y(z+a)[u(z-a) - u(z+a)] \quad (17)$$

From the probability theory, we know that for $N$ independent random variables, the pdf of their sum is the convolution of their pdfs [8]. Hence, we can write:

$$f_{Z|H_0}(z) = \underbrace{f_{z_{i|H_0}}(z) * \ldots * f_{z_{i|H_0}}(z)}_{N}. \quad (18)$$

As source coding suggests, by assuming the probability of 0 and 1 to be equal, error probability becomes:

$$P_{err} = \frac{1}{2} P_{err|H_0} + \frac{1}{2} P_{err|H_1}, \quad (19)$$

Due to the symmetric property of the problem, we already know that the error probability conditioned on $H_0$ or $H_1$ are equal and this error probability can be approximated with some mathematical manipulations as follows:

$$P_{err|H_0} = p(Z > 0 | H_0) = \int_{0}^{+\infty} f_{Z|H_0}(z) dz$$
$$\approx \left(\frac{1}{1+2P}\right)^N \sum_{i=\frac{N}{2}+1}^{N} \binom{N}{i} P^{N-i}, \quad (20)$$

and

$$P_{err} \approx \left(\frac{1}{1+2P}\right)^N \sum_{i=\frac{N}{2}+1}^{N} \binom{N}{i} P^{N-i}. \quad (21)$$

This error probability is a function of *SNR*, *a*, and *N*, which one can achieve a desired error probability by suitably choosing these parameters.

V. SIMULATION RESTULTS

Simulation results are shown in this section. Simulations are done on H.264 standard software reference JM 18.0 [9]. We select Carphone sequence ($176 \times 144$) in QCIF format with frame rate of $30^{fps}$. We simulated both optimum and sub-optimum decoders.

Figure 3 shows the BER plots of receivers.

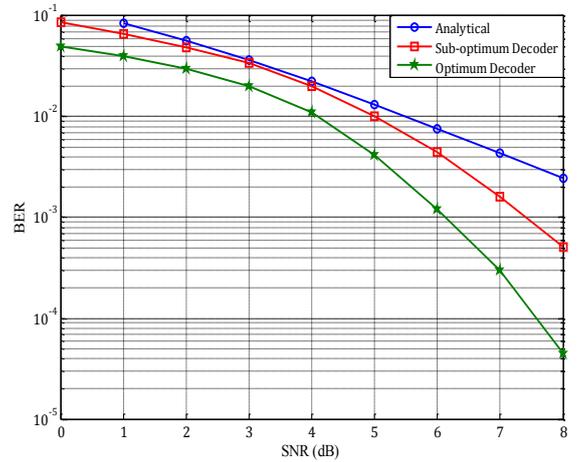

Figure 3. BER vs. SNR for optimum and sub-optimum decoders (*N=120*, *a=1*).

For different values of *N* we have run the simulation and the results are depicted in Figure 4. As expected, by increasing the value of *N*, we see that the BER decreases.

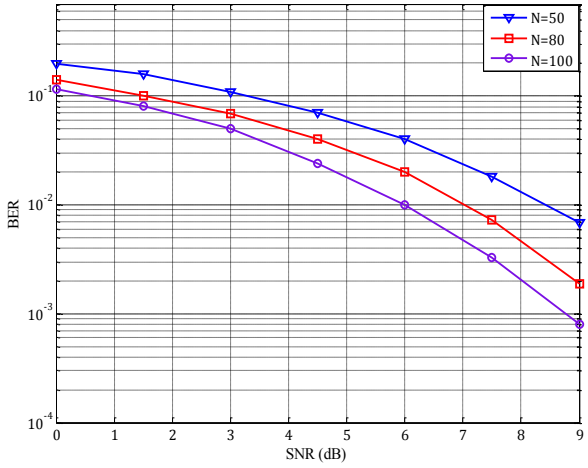

Figure 4. BER vs. SNR for optimum decoder for different values of *N* for Carphone sequence.

Simulations have also been run for Foreman and Akiyo sequences and the results are depicted in Figure 5 and 6, respectively.

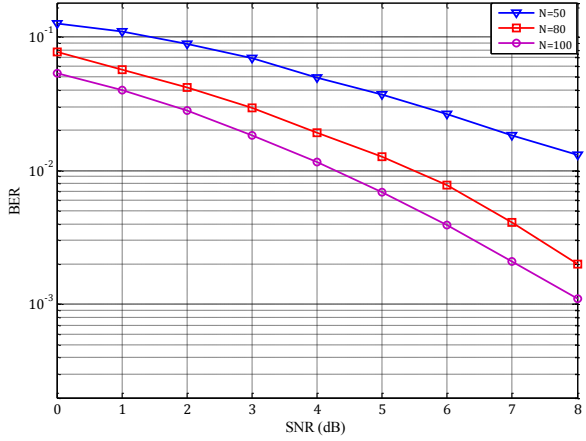

Figure 5. BER vs. SNR for optimum decoder for different values of *N* for Foreman sequence.

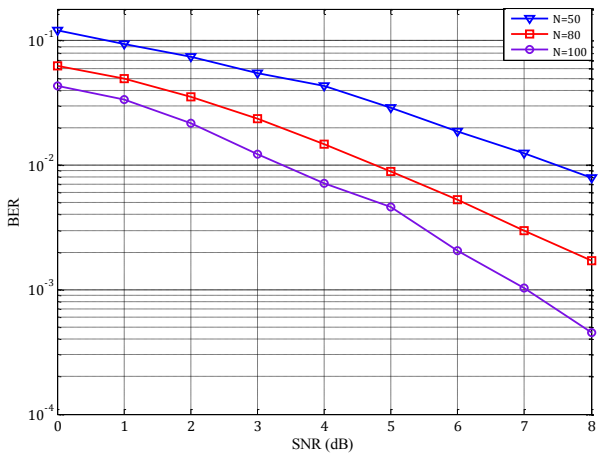

Figure 6. BER vs. SNR for optimum decoder for different values of *N* for Akiyo sequence

As it is clear from the figures above, watermarking in the Akiyo sequence has better performance than the Foreman and Carphone sequences. This is due to the fact that the Akiyo sequence can be better modeled with the Laplace distribution.

In Figures 7 and 8, the first frame of the Foreman and Akiyo sequences are shown, respectively.

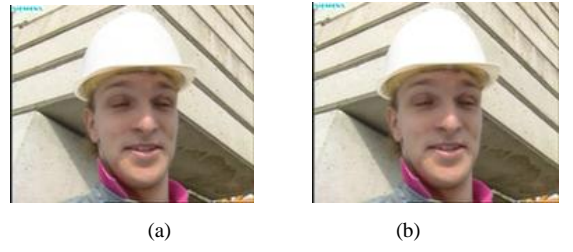

(a)      (b)

Figure 7. First Frame of the Foreman sequence. (a) Original frame. (b) Watermarked frame.

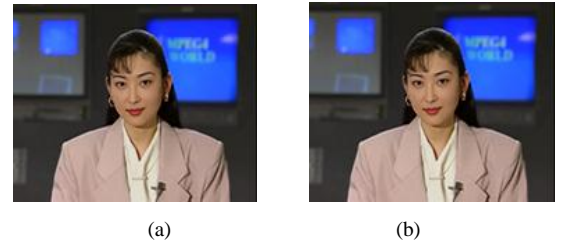

(a)      (b)

Figure 8. First Frame of the Akiyo sequence. (a) Original frame. (b) Watermarked frame.

We have also investigated the perceptual quality of the mentioned watermarking scheme with both the peak signal-to-noise ratio (PSNR) and the video quality metric (VQM) [10]. VQM is a metric that shows the impairment degree of the watermarked sequence and takes the values between zero and one, for which zero represents no distortion and one represents maximum impairment level. For the Foreman and Akiyo sequences the values of PSNR are 42.41 dB and 44.24 dB, and the values of VQM are 0.023 and 0.031, respectively. We can see that this watermarking scheme has suitable transparency.

In Table I, the results for the bitrate increase due to watermarking is reported for different video sequences.

TABLE I. BITRATE INCREASE FOR DIFFERENT VIDEO SEQUENCES

| Video Sequence | Bitrate Increase |
|---|---|
| Foreman | 0.47% |
| Akiyo | 0.31% |
| Carphone | 0.35% |
| Mobile | 0.48% |

The authors in [11] used the Cauchy distribution to model the DCT coefficients and proposed an additive watermarking method. In the same watermarking capacity and for the Foreman and Akiyo sequences, our bitrate increase was less than their simulation results. For example, our bitrate increase for the Akiyo and Foreman sequences are 0.31% and 0.47%, respectively, whereas in [11], these values were 1.92% and 1.55%.

## VI. Conclusion

An additive watermarking method with Laplacian noise was investigated. By analyzing the optimum decoder and simplifying the decision rule, we found the error probability for the sub-optimum decoder. Lowering the complexity order of the decoder was shown to be reasonable, since the performance of the sub-optimum decoder was close to the optimum one, as demonstrated in the simulation part. Hence, we have also showed that the BER decreases as the number of selected DCT coefficients, $N$, increases. Our watermarking scheme was shown to have enough transparency and acceptable bitrate increase.